\documentclass{article}
\usepackage{amsfonts}
\usepackage{amssymb,amsmath,amsthm,latexsym}
\newtheorem{theorem}{Theorem}[section]
\newtheorem{corollary}[theorem]{Corollary}
\newtheorem{proposition}[theorem]{Proposition}
\newtheorem{lemma}[theorem]{Lemma}

\newtheorem{example}[theorem]{Example}

\newtheorem{remark}[theorem]{Remark}
\numberwithin{equation}{section}

\begin{document}

\title{Homogeneous Weights of Matrix Product Codes
over Finite Principal Ideal Rings}

\author{Yun Fan${}^1$, San Ling${}^2$, Hongwei Liu${}^1$}
\date{${}^1$School of Mathematics and Statistics,
 Central China Normal University, Wuhan 430079, China \\
${}^2$Division of Mathematical Sciences, School of Physical \& Mathematical Sciences,
         Nanyang Technological University, Singapore 637616, Singapore}

\maketitle

\begin{abstract}

In this paper, the homogeneous weights of matrix product codes
over finite principal ideal rings are studied and a lower bound
for the minimum homogeneous weights of such matrix product codes
is obtained.


\end{abstract}

\section{Introduction}

Matrix product codes over finite fields were introduced in \cite{BN}.
Many well-known constructions can be formulated as matrix product codes,
for example, the $(a|a+b)$-construction, the $(a+x|b+x|a+b+x)$-construction,
and, some quasi-cyclic codes can be rewritten as matrix product codes,
see \cite{LS}. The reference \cite{BN} also introduced non-singular by
columns matrices and exhibited a lower bound for the minimum Hamming distances
of matrix product codes over finite fields constructed by such matrices.
More references on matrix product codes appeared later, e.g., in
\cite{HLR,HR10,HR11,MS,OS}.

Codes over finite rings have also been studied from many
perspectives since the seminal work \cite{HKCSS}.
It was also shown later in \cite{W-99} that a finite Frobenius ring is
suitable as an alphabet for linear coding. Further,
\cite{GNW} showed that, for any finite ring, there is a Frobenius module
which is suitable as an alphabet for linear coding.
Inspired by the idea of module coding,
\cite{W-08} proved that the biggest class of finite rings which
are suitable as alphabets for linear coding consists
of the finite Frobenius rings.

Finite principal ideal rings form an important subclass of finite
Frobenius rings. In particular, all the residue rings ${\bf Z}_N$
of integers modulo an integer $N>1$ are principal ideal rings.
It is well known that a finite commutative ring is a principal ideal ring
if and only if it is a product of finite chain rings.
The reference \cite{Van-Asch} extended the lower bound obtained in \cite{BN}
for the minimum Hamming distances of matrix product codes with non-singular by
columns matrices over finite fields to the minimum homogeneous
weights of matrix product codes over finite chain rings.

In this paper, we consider matrix product codes
over finite commutative principal ideal rings, and extend the
result on the lower bound for the minimum homogeneous weights
of matrix product codes over finite chain rings
to matrix product codes over finite commutative principal ideal rings.

In the next section, necessary notations and fundamentals
are introduced as preliminaries. In Section 3,
we state our main theorem, its corollaries and some remarks.
Since the proof of the main theorem is long and technical,
it is deferred to Section 4.

\section{Preliminaries}

In this paper, $R$ is always a finite commutative ring.

For the finite commutative ring $R$ and a positive integer $\ell$,
any non-empty subset $C$ of $R^\ell$
is called a code over $R$ of length $\ell$,
or more precisely, an $(\ell,M)$ code over $R$,
where $M=|C|$ denotes the cardinality of $C$;
the code $C$ over $R$ is said to be linear
if $C$ is an $R$-submodule of $R^\ell$.
Recall that the usual Hamming weight $w_H$ on $R$, i.e.,
$w_H(0)=0$ and $w_H(r)=1$ for all non-zero $r\in R$,
induces in a standard way the {\em Hamming weight} on $R^\ell$,
denoted by $w_H$ again, and the {\em Hamming distance} $d_H$ on $R^\ell$
as follows: $w_H({\bf x})=\sum\limits_{i=1}^\ell w_H(x_i)$ for
${\bf x}=(x_1,\cdots,x_\ell)\in R^\ell$,
and $d_H({\bf x},{\bf x}')=w_H({\bf x}-{\bf x}')$ for
${\bf x},{\bf x}'\in R^\ell$.
We also let $d_H(C)=\min\limits_{{\bf c}\ne{\bf c}'\in C} d_H({\bf c},{\bf c}')$.
This is known as the {\em minimum Hamming distance} of the code $C$.

On the other hand,
a {\em homogeneous weight} on the finite commutative ring $R$
is defined to be a non-negative real function
$w_h$ from $R$ to the real number field which satisfies the following two
conditions:

\begin{itemize}
\item $w_h(r)=w_h(r')$ for $r,r'\in R$, provided $Rr=Rr'$,

\item there is a positive real number $\lambda$
such that $\sum\limits_{x\in Rr}w_h(x)=\lambda|Rr|$
for any non-zero $r\in R$,
where $|Rr|$ denotes the cardinality of the set $Rr$.
\end{itemize}

\noindent
It has been shown in \cite{GS}
that such a function is uniquely determined (up to a scalar $\lambda$)
on $R$ as follows
\begin{equation}\label{eq2.1}
 w_h(r)=\lambda\left(1-\frac{\mu(0,Rr)}{\varphi(Rr)}\right),
\end{equation}
where $\mu$ is the M\"obius function on the lattice of
all the principal ideals of $R$, and $\varphi(Rr)$
denotes the number of elements $x\in Rr$ such that $Rx=Rr$.
Thus, the homogeneous weight is uniquely determined
up to a positive multiple $\lambda$. In the rest of the paper,
we always take $\lambda=1$ in (\ref{eq2.1}) for convenience,
and denote the uniquely determined homogeneous weight by $w_h$.
As with the Hamming weight,
the function $w_h$ on the ring $R$ induces a function $w_h$ on $R^\ell$
and a two-variable function $d_h$ on $R^\ell$; and
$d_h(C)=\min\limits_{{\bf c}\ne{\bf c}'\in C} d_h({\bf c},{\bf c}')$
is said to be the {\em minimum homogeneous distance} of the code $C$.

Let $A=(a_{ij})_{m\times \ell}$ be an $m\times\ell$ matrix over
the finite commutative ring $R$, and let
$C_1,\cdots,C_m$ be codes over $R$ of length $n$. Then
$$
 C=[C_1,\cdots,C_m]A=\big\{({\bf c}_1,\cdots,{\bf c}_m)A
  \mid {\bf c}_1\in C_1,\cdots,{\bf c}_m\in C_m\big\}
$$
is called a {\em matrix product code}, where the codewords
${\bf c}_j$ of $C_j$ are written as column vectors, hence
$({\bf c}_1,\cdots,{\bf c}_m)$ are $n\times m$ matrices.

We say that a square matrix over $R$ is {\em non-singular} if
its determinant is a unit of $R$. By usual linear algebra,
a non-singular matrix over $R$ is an invertible matrix over $R$.
Following \cite{BN}, we say that the $m \times\ell$ matrix $A$ is
{\em non-singular by columns} if, for any $k\le m$, any
$k\times k$ determinant within the first $k$ rows of $A$ is a unit of $R$.
It is clear that, if $A$ is non-singular by columns,
then any matrix obtained from $A$ by permuting its columns is still
non-singular by columns.
We say that a matrix $A$ is {\it column-permutably upper triangular}
if $A$ can be transformed by some suitable permutation of the columns
to an upper triangular matrix $A'=(a'_{ij})_{m\times\ell}$
(i.e., $a'_{ij}=0$ for all $1\le j<i\le m$).

\medskip
From now on, we always assume that $R$ is a finite commutative
{\em principal ideal ring}, i.e., $R$ is a finite commutative ring in
which any ideal can be generated by one element, or equivalently,
there are finite chain rings $R_1$, $\cdots$, $R_s$
and an isomorphism:
\begin{equation}\label{PIR as product}
 R~\mathop{\longrightarrow}^{\cong}~R_1\times\cdots\times R_s\,,\qquad
 r\longmapsto(r_1,\cdots,r_s).
\end{equation}
With this isomorphism, we can identify
$R$ with $R_1\times\cdots\times R_s$ and write $r=(r_1,\cdots,r_s)$.
For $t=1,\cdots,s$, by $J_t$ we denote the unique maximal ideal
of the chain ring $R_t$ (note that $J_t=0$ if $R_t$ is a field).
Hence $R_t/J_t$ is a finite field, and we further assume that
\begin{equation}\label{head of PIR}
F_t=R_t/J_t\cong{\rm GF}(q_t)~~\mbox{for}~ t=1,\cdots,s,\quad
 \mbox{and}~~ q_1\le q_2\le\cdots\le q_s,
\end{equation}
where ${\rm GF}(q_t)$ denotes the Galois field of order $q_t$.
For each $t$, there is an integer $e_t \ge 1$,
called the {\em nilpotency index of the chain ring $R_t$}, such that
\begin{equation}\label{nil of PIR}
J_t^{e_t-1}\ne 0\quad{\rm but}\quad J_t^{e_t}=0,\qquad t=1,\cdots,s.
\end{equation}
Note that $J_t^0=R_t$ for any $t$.

\medskip
We list some easy facts for later use.
Since any ideal $I$ of $R$ has the form
$I=I_1\times\cdots\times I_s$ with $I_t$ being an
ideal of $R_t$ for $t=1,\cdots,s$, it follows that
$R/I=R_1/I_1\times\cdots\times R_s/I_s$ is still
a principal ideal ring.

\medskip
Next, if elements $u_{t1}, u_{t2}, \cdots, u_{tq_t}$ of $R_t$ satisfy that
$$F_t=R_t/J_t=\big\{u_{t1}+J_t,\,u_{t2}+J_t,\,\cdots,\,u_{tq_t}+J_t\big\},$$
then, for any integer $k$ with  $0<k\le e_t$ and any element $a$ of
the set difference $J_t^{k-1} \setminus J_t^{k}$, we have that
\begin{equation}\label{steps PIR}
J_t^{k-1}/J_t^{k}=\big\{u_{t1}a+J_t^{k},\;u_{t2}a+J_t^{k},\;
  \cdots,\;u_{tq_t}a+J_t^{k}\big\}.
\end{equation}

\medskip
Recall Formula (\ref{eq2.1}) and rewrite it as (recall that we have set $\lambda=1$):
$$\textstyle
 w_h(r)=1-\frac{\mu(r)}{\varphi(r)},
$$
where $\mu(r)=\mu(0,Rr)$ and $\varphi(r)=\varphi(Rr)$.
Since $R$ is a product of chain rings as in Eqn (\ref{PIR as product}) and
$(r_1,\cdots,r_s)\in R_1\times\cdots\times R_s$,
both $\mu$ and $\varphi$ satisfy that
$$
 \mu\big((r_1,\cdots,r_s)\big)=\mu(r_1)\cdots\mu(r_s),\quad
 \varphi\big((r_1,\cdots,r_s)\big)=\varphi(r_1)\cdots\varphi(r_s);
$$
thus, the homogeneous weight $w_h$ on $R$ is (see \cite[Theorem 4.1]{FL})
$$\textstyle
 w_h(r)=w_{h}(r_1,\cdots,r_s)
 =1-\prod_{t=1}^s\frac{\mu(r_t)}{\varphi(r_t)}.
$$
Further, for $r_t\in R_t$,
there is (as long as $r_t\ne 0$) a unique integer $f_{r_t}$ with $0<f_{r_t}\le e_t$ such that
$r_t\in J_t^{e_t-f_{r_t}} \setminus J_t^{e_t-f_{r_t}+1}$, then we have
(see \cite{FL} for details):
$$
 \mu(r_t)=\begin{cases}1,&r_t=0,\\-1,& f_{r_t}=1,\\0,&f_{r_t}>1,\end{cases} \qquad\qquad
 \varphi(r_t)=\begin{cases}1, & r_t=0,\\ q_t-1, & f_{r_t}=1,\\
  q_t^{f_{r_t}}-q_t^{f_{r_t}-1}, & f_{r_t}>1.\end{cases}
$$
For a non-zero element $r=(r_1,\cdots,r_s)$ of $R$, setting
\begin{equation}\label{eq2.6}
T_r=\{1\le t\le s \mid r_t\ne 0\},\qquad
  \bar T_r=\{t\in T_r\mid r_t\in J_t^{e_t-1}\},
\end{equation}
we obtain a formula to calculate the homogeneous weight on $R$ as follows:
\begin{equation}\label{eq2.7}
 w_{h}(r_1,\cdots,r_s)=\begin{cases}0, & r=0,\\ 1, & \bar T_r\ne T_r,\\
    1-(-1)^{|T_r|}\prod\limits_{t\in T_r}\frac{1}{q_t-1},& \bar T_r=T_r.
  \end{cases}
\end{equation}
From Formula (\ref{eq2.7}) one can see that (take any $q_2\ge q_1$ if $s=1$):
\begin{equation}\label{eq2.8}
\textstyle
  1-\frac{1}{(q_1-1)(q_2-1)}\le w_h(r)\le 1+\frac{1}{q_1-1},\qquad
   \forall~~ 0\ne r\in R.
\end{equation}

\medskip
We next recall a few
facts on matrices over a finite commutative principal ideal ring $R$.
For any $t$, by (\ref{head of PIR}), we have a surjective homomorphism
\begin{equation}\label{eq2.9}
\rho_t:~ R\longrightarrow F_t,\qquad r\longmapsto\rho_t(r)
\end{equation}
with kernel
$I_t=R_1\times\cdots\times R_{t-1}\times J_t\times R_{t+1}\times\cdots\times R_s$,
i.e., $R/I_t\cong R_t/J_t=F_t$.
By a fundamental argument on determinants in linear algebra,
one can prove (alternatively, a proof may be found in classical references such as \cite{M}):

\begin{lemma}\label{lemma 2.1}
Let $A=(a_{ij})_{m\times \ell}$ be an $m\times \ell$ matrix over $R$.
\begin{itemize}
\item[(i)] If $A$ is non-singular by columns, then,
for any non-trivial quotient ring $\bar R=R/I$
 (i.e., $I$ is an ideal of $R$ with $I\ne R$),
the matrix $\bar A=\big(\bar a_{ij}\big)_{m\times \ell}$ over $\bar R$
is non-singular by columns.

\item[(ii)] If the matrix $\rho_t(A)=\big(\rho_t(a_{ij})\big)_{m\times\ell}$
over $F_t$ is non-singular by columns for all $t=1,\cdots,s$,
then $A$ is non-singular by columns.
\end{itemize}
\end{lemma}

By the above lemma and with the help of Eqn (\ref{steps PIR}),
it is easy to prove the following result which is an extension of
\cite[Prop. 3.3]{BN} and \cite[Prop. 1]{Van-Asch}.

\begin{lemma}\label{lemma 2.2}
Assume that $m>1$. There exists an $m\times\ell$ matrix over $R$
which is non-singular by columns
if and only if $m\le \ell\le\min\{q_1,\cdots,q_s\}$.
\end{lemma}

The following result has appeared in \cite[Lemma 4.1]{FLL}.

\begin{lemma}\label{lemma 2.3}
Assume that an $m\times\ell$ matrix $A$ over $R$
is non-singular by columns and $1\le k\le m$.
Then the minimum Hamming distance of the linear code in $R^\ell$
generated by the first $k$ rows of $A$ is $\ell-k+1$.
\end{lemma}

\section{The main results}

We keep the notations of (\ref{PIR as product}),
(\ref{head of PIR}) and (\ref{nil of PIR}).
In this section, we state our main theorem, its corollaries and
some remarks. The main theorem will be proved in the next section.

\begin{theorem}\label{Theorem}
Let the notations be as in (\ref{PIR as product}) and (\ref{head of PIR}),
and assume that $q_2>q_1+1$ provided $s>1$.
Let $A=(a_{ij})_{m\times\ell}$ be an $m\times\ell$ matrix over $R$
which is non-singular by columns, and let
$C_j$ be an $(n,M_j)$ code over $R$, for $j=1,\cdots,m$.
Then $C=[C_1,\cdots,C_m]A$ is an
$\big(n\ell,\prod\limits_{j=1}^m M_j\big)$-code over $R$ with
\begin{equation}\label{main ineq}
 d_h(C)\ge\min\big\{\ell d_h(C_1),\;(\ell-1)d_h(C_2),\;
      \cdots,\;(\ell-m+1)d_h(C_m)\big\}.
\end{equation}
Furthermore, equality holds in (\ref{main ineq})
if one of the following conditions is satisfied:
\begin{itemize}
\item[{\bf(C1)}] $A$ is column-permutably upper triangular;

\item[{\bf(C2)}] $C_1,C_2,\cdots,C_m$ are linear codes and $C_1\supseteq
C_2\supseteq\cdots\supseteq C_m$.
\end{itemize}
\end{theorem}

\medskip
With the help of the results in \cite{FLL}, we have a consequence
of the theorem for the dual codes of matrix product codes.

\begin{corollary}\label{cor for dual}
Keep the notations as in (\ref{PIR as product}), (\ref{head of PIR}),
and assume that $q_2>q_1+1$ provided $s>1$.
If $A$ is an $m\times m$ matrix over $R$ which is
non-singular by columns, and $C_j$ is an $(n,M_j)$-linear code over
$R$, for $j=1,\cdots,m$, then the dual code $C^\bot$ of the matrix
product code $C=[C_1,\cdots,C_m]A$ is an
$\big(nm,~\prod\limits_{j=1}^m(|R|^n/M_j)\big)$-linear code over
$R$ with
$$
 d_h(C^\bot)\ge\min\big\{md_h(C_m^\bot),\;(m-1)d_h(C_{m-1}^\bot),\;
    \cdots,\;1\cdot d_h(C_1^\bot)\big\}.
$$
Furthermore, equality holds if one of the following conditions is satisfied:
\begin{itemize}
\item[{\bf(C1)}] $A$ is column-permutably upper triangular;

\item[{\bf(C2)}] $C_1,C_2,\cdots,C_m$ are linear codes and
$C_1\supseteq C_2\supseteq\cdots\supseteq C_m$.
\end{itemize}
\end{corollary}

\smallskip\noindent{\bf Proof.}~  For a square matrix $A$ over $R$ which
is non-singular by columns, it is shown in \cite[Theorem 3.3]{FLL} that
$A$ is invertible and $J(A^{-1})^T$ is non-singular by columns
too, where $(A^{-1})^T$ denotes the transpose of the inverse $A^{-1}$ and
$$J=\begin{pmatrix}0&\cdots&0&1\\0&\cdots&1&0\\
  \vdots&\vdots&\vdots&\vdots\\1&\cdots&0&0\\\end{pmatrix},$$
and $C^\bot=[C_1^\bot,\cdots,C_m^\bot](A^{-1})^T$.
Noting that $JJ=I$, where $I$ denotes the identity matrix,
and $[C_1^\bot,\cdots,C_m^\bot]J=[C_m^\bot,\cdots,C_1^\bot]$, we have that
$$
 C^\bot=[C_1^\bot,\cdots,C_m^\bot]JJ(A^{-1})^T=[C_m^\bot,\cdots,C_1^\bot]J(A^{-1})^T.
$$
It is easy to check that, if $A$ satisfies (C1) then
so does $J(A^{-1})^T$; and similarly for (C2).
Thus the conclusions are derived from Theorem \ref{Theorem}. \qed

\medskip In fact, in \cite{FLL},
a very precise description for the structure of
$C^\bot$, where $C=[C_1,\cdots,C_m]A$, was obtained in a more general
setting, where $R$ is any finite commutative Frobenius ring and
$A$ does not need to be square and non-singular by columns.
It is therefore possible to obtain a lower bound for the
minimum homogeneous distance of $C^\bot$ in that general setting,
see \cite{FLL-1}.

\begin{remark}\rm
In the case when $s=1$, i.e., $R$ is a finite chain ring,
Theorem \ref{Theorem} contains the result \cite[Proposition 2]{Van-Asch}
and a generalization of the main result of~\cite{HLR};
moreover, Corollary \ref{cor for dual} bounds from below
the homogeneous distance of the dual codes of matrix product codes.
\end{remark}

\smallskip
The residue ring ${\bf Z}_N$ of integers modulo an integer $N>1$
is one of the best-known finite principal ideal rings.
Writing $N=p_1^{e_1}\cdots p_s^{e_s}$, where $p_1<\cdots<p_s$ are
primes and $e_t>0$ for $t=1,\cdots,s$,
we see that the Chinese Remainder Theorem:
\begin{equation*}
\begin{array}{ccc}
 {\bf Z}_N &\cong& {\bf Z}_{p_1^{e_1}}
  ~~\times~~\cdots~~\times~~{\bf Z}_{p_s^{e_s}}\,,\\
 r~({\rm mod}~N)&\mapsto&\big(r~({\rm mod}~p_1^{e_1}),\;\cdots,\;
   r~({\rm mod}~p_s^{e_s})\big),
\end{array}
\end{equation*}
is just the version for ${\bf Z}_N$ of the decomposition (\ref{PIR as product}).
Therefore, the assumption ``$q_2>q_1+1$ provided $s>1$'' in
Theorem \ref{Theorem} translates into the assumption
``$p_2\ne 3$ provided $p_1=2$'' for ${\bf Z}_N$; and
we obtain the following result from Theorem \ref{Theorem} at once.

\begin{corollary}\label{Z_N}
Let $N>1$ be an integer which is not divisible by $6$. Let $A$ be an
$m\times \ell$ matrix over ${\bf Z}_N$ which is non-singular by
columns, and let $C_j$ be an $(n,M_j)$-code over ${\bf Z}_N$, for
$j=1,\cdots,m$. Then $C=[C_1,\cdots,C_m]A$ is an
$\big(n\ell,\prod\limits_{j=1}^m M_j\big)$-code over ${\bf Z}_N$ with
\begin{equation}\label{Ineq Z_N}
 d_h(C)\ge\min\big\{\ell d_h(C_1),\;(\ell-1)d_h(C_2),\;
   \cdots,\;(\ell-m+1)d_h(C_m)\big\}.
\end{equation}
Furthermore, equality holds if one of the following conditions is satisfied:
\begin{itemize}
\item[{\bf(C1)}] $A$ is column-permutably upper triangular;

\item[{\bf(C2)}] $C_1,C_2,\cdots,C_m$ are linear codes and $C_1\supseteq
C_2\supseteq\cdots\supseteq C_m$.
\end{itemize}
\end{corollary}

There is also an analogous version of Corollary \ref{cor for dual} for ${\bf Z}_N$,
with the same assumption ``$N$ is not divisible by $6$''.

\begin{remark}\rm
Recall that, to be a geometric distance,
a two-variable real function must meet three conditions:
it is positive, it is symmetric, and it satisfies the triangle inequality.
It is known that the homogeneous distance $d_h$ may not be
a geometric distance.
References \cite{CH} and \cite{HN} contain extensive studies on
weights on integral residue rings: in particular,
a necessary and sufficient condition
for the homogeneous distance $d_h$ on ${\bf Z}_N^\ell$ to be
a geometric distance is that $N$ is not divisible by $6$.
By Corollary \ref{Z_N} and Example \ref{ex3.1} below,
this condition is also necessary and sufficient for
Inequality (\ref{Ineq Z_N}) to hold.
\end{remark}

The assumption ``$q_2>q_1+1$ provided $s>1$''
in Theorem \ref{Theorem} will play a crucial role in the proof
of the theorem. Moreover, the following example illustrates
that the assumption cannot be removed.

\begin{example}\label{ex3.1}
{\rm
Let the notations be as in (\ref{PIR as product}),
(\ref{head of PIR}) and (\ref{nil of PIR}).
Assume that $s>1$ and $q_1+1=q_2\le q_3\le\cdots\le q_s$ and let $q=q_1$.
Let $u_{t1},u_{t2},\cdots,u_{tq_t}\in R_t$ be as in Eqn (\ref{steps PIR}).
In the present case, we can choose them as follows:
\begin{itemize}
\item for $t=1$, $u_{11}+J_1,\;\cdots,u_{1q}+J_1$ are just all the elements
of $F_1=R_1/J_1$;

\item for $t=2$, since $q_2=q+1$, we can take
$u_{21}+J_2,\;\cdots,\;u_{2q}+J_2$ to be all non-zero elements
of $F_2=R_2/J_2$;

\item for $t\ge 3$, since $q_t>q$, we can take
$u_{t1}+J_t,\;\cdots,\;u_{tq}+J_t$ to be distinct elements of $F_t=R_t/J_t$.
\end{itemize}

\noindent Let
$\beta_j=(u_{1j},u_{2j},\cdots,u_{sj})\in R
  =R_1\times\cdots\times R_s$ for $j=1,\cdots,q$, and let
$$A=\begin{pmatrix}1&1&\cdots&1\\ \beta_1&\beta_2&\cdots&\beta_q\end{pmatrix}.$$
It is easy to see (cf. Lemma \ref{lemma 2.2}) that
$A$ is non-singular by columns. Let
$$a=(a_1,0,0,\cdots,0),\; b=(0,b_2,0,\cdots,0) \in R=R_1\times\cdots\times R_s , $$
where $a_1\in J_1^{e_1-1} \setminus \{0\}$ and $b_2\in J_2^{e_2-1} \setminus \{0\}$.
Set $C_1=Ra$ and $C_2=Rb$, then both are linear codes over $R$ of length $1$.
Then we have the matrix product code $C=[C_1,C_2]A$.
By Formula (\ref{eq2.7}), we have that
$$\textstyle
 d_h(C_1)=w_h(a)=1+\frac{1}{q-1}\,,\quad
 d_h(C_2)=w_h(b)=1+\frac{1}{q_2-1}=1+\frac{1}{q}.
$$
Since $1+\frac{1}{q-1}>1+\frac{1}{q}$, we get
$$\textstyle
\min\{qd_h(C_1),\,(q-1)d_h(C_2)\}=(q-1)\big(1+\frac{1}{q}\big)=q-\frac{1}{q}.
$$
On the other hand, there is a codeword ${\bf c}$ of $C$ as follows:
$$
 {\bf c}=(a,b)A=(a+b\beta_1,~a+b\beta_2,~\cdots,~a+b\beta_q)
$$
with
$$
 a+b\beta_j=(a_1,\;b_2 u_{2j},\;0,\cdots,0),\,\qquad j=1,\cdots,q.
$$
By Eqn (\ref{steps PIR}),
$b_2 u_{2j}$, for $j=1,\cdots, q$, are all the non-zero elements of $J_2^{e_2-1}$,
and by Formula (\ref{eq2.7}),
$w_h(a+b\beta_j)=1-\frac{1}{(q_1-1)(q_2-1)}=1-\frac{1}{q(q-1)}$, so
$$\textstyle
 w_h({\bf c})=q\big(1-\frac{1}{q(q-1)}\big)=q-\frac{1}{q-1}
 <q-\frac{1}{q}=\min\{qd_h(C_1),\,(q-1)d_h(C_2)\}.
$$
Therefore,
$$
 d_h(C)<\min\{qd_h(C_1),\,(q-1)d_h(C_2)\},
$$
which implies that Inequality (\ref{main ineq}) does not hold
for the matrix product code $C=[C_1,C_2]A$.
}
\end{example}

\section{Proof of Theorem \ref{Theorem}}

We continue to keep the notations of (\ref{PIR as product}), (\ref{head of PIR})
and (\ref{nil of PIR}) and assume that $q=q_1\le q_2\le\cdots\le q_s$.

Let $A=(a_{ij})_{m\times \ell}$ be a matrix over $R$
which is non-singular by columns;
then $\ell\le q=q_1$ if $m>1$ (see Lemma \ref{lemma 2.2}).
Let $C_1,\cdots,C_m$ be codes over $R$ of length~$n$.
Consider the matrix product code
\begin{equation}\label{eq4.1}
 C=[C_1,\cdots,C_m]A=\left\{({\bf c}_1,\cdots,{\bf c}_m)A\,|\,
  {\bf c}_1\in C_1,\, \cdots,\,{\bf c}_m\in C_m \right\}.
\end{equation}

Since the proof of Inequality (\ref{main ineq}) is long and delicate,
we put the key steps in Subsections 4.1--4.3; these subsections
show that the following inequality holds for all $1 \le k \le m$ by
splitting into various cases:
$$w_{h}\big(({\bf c}_1,\cdots,{\bf c}_k,{\bf 0},\cdots,{\bf 0})A\big)
  \ge(\ell-k+1)w_h({\bf c}_k).$$
Subsection 4.4 then completes the proof of Theorem \ref{Theorem}.

\subsection{$k\times\ell$ Non-singular by Columns Matrices}
Let $A$ be as above, let $2\le k\le m$,
and let $A_1$, $\cdots$, $A_k$ be the first $k$ rows of $A$.

Let $r_1,\cdots,r_k\in R$ with $r_k\ne 0$
and let $\alpha=r_1A_1+\cdots+r_kA_k\in R^\ell$.
We have seen from Lemma \ref{lemma 2.3} that
$$w_H(\alpha)=w_H(r_1A_1+\cdots+r_kA_k)\ge \ell-k+1.$$

\begin{lemma}\label{lem4.1}
Let the notations be as above.
\begin{enumerate}
\item[(i)] If $w_H(r_1A_1+\cdots+r_kA_k)=\ell-k+1$, then
$w_h(r_1A_1+\cdots+r_kA_k)=(\ell-k+1)w_h(r_k)$.

\item[(ii)] If $w_H(r_1A_1+\cdots+r_kA_k)>\ell-k+1$
and one of the following two conditions holds:
\begin{itemize}
\item $k\ge 3$,

\item $k=2$ and $\ell<q_1$,
\end{itemize}
\end{enumerate}
then $w_h(r_1A_1+\cdots+r_kA_k)\ge(\ell-k+1)w_h(r_k)$.
\end{lemma}

\noindent{\bf Proof.}~ Write
$\alpha=r_1A_1+\cdots+r_kA_k=(\alpha_1,\cdots,\alpha_\ell)$, where
$$\alpha_j=r_1a_{1j}+\cdots+r_ka_{kj},\qquad j=1,\cdots,\ell.$$

(i)~ Since $w_H(r_1A_1+\cdots+r_kA_k)=\ell-k+1$,
there are exact $k-1$ zeros among $\alpha_1,\cdots,\alpha_\ell$.
Without loss of generality, we assume that $\alpha_1=\cdots=\alpha_{k-1}=0$
and $\alpha_j\ne 0$, for $j=k,k+1,\cdots,\ell$. In particular,
$$\left\{\begin{array}{ccc} r_1a_{11}+\cdots+r_{k-1}a_{k-1,1}&=&-r_ka_{k1}\\
r_1a_{12}+\cdots+r_{k-1}a_{k-1,2}&=&-r_ka_{k2}\\\vdots\quad\vdots\quad\vdots&&\vdots\\
r_1a_{1,k-1}+\cdots+r_{k-1}a_{k-1,k-1}&=&-r_ka_{k,k-1}.
\end{array}\right.$$
By the non-singularity by columns of $A$, the coefficient matrix
$(a_{ji})_{(k-1)\times(k-1)}$ of the above linear system is
non-singular, i.e., invertible, hence $r_1,\cdots,r_{k-1}$ are all
linear combinations of $r_ka_{k1},\cdots,r_ka_{k,k-1}$. Therefore, $r_j\in
Rr_k$ for $j=1,\cdots,k-1,k$; hence all $\alpha_j\in Rr_k$, i.e.,
$$
 0\ne R\alpha_j\subseteq Rr_k\,,\qquad j=k,k+1,\cdots,\ell.
$$
Suppose that there is an index $t$ with $k\le t\le \ell$
such that $R\alpha_t\subsetneqq Rr_k$.
Considering the quotient ring $\bar R=R/R\alpha_t$,
then $\bar r_k\ne 0$ and $\bar A=(\bar a_{ij})_{m\times \ell}$
is still non-singular by columns (see Lemma \ref{lemma 2.1}). However,
$$\bar\alpha_1=\cdots=\bar\alpha_{k-1}=0\quad{\rm and}\quad \bar\alpha_t=0, $$
hence
$$ w_H(\bar r_1\bar A_1+\cdots+\bar r_k\bar A_k)\le \ell-k<\ell-k+1, $$
which contradicts the fact that
$w_H(\bar r_1\bar A_1+\cdots+\bar r_k\bar A_k)\ge \ell-k+1$ (see Lemma \ref{lemma 2.3}).
Therefore, $R\alpha_j=Rr_k$ for all $j=k,k+1,\cdots,\ell$, and
$$ w_h(r_1A_1+\cdots+r_kA_k)=\sum_{j=k}^\ell w_h(\alpha_j)
  =\sum_{j=k}^\ell w_h(r_k)=(\ell-k+1)w_h(r_k). $$

(ii)~ In this case, there are at least $\ell-k+2$ non-zeros
among $\alpha_1,\cdots,\alpha_\ell$. By Inequality (\ref{eq2.8}), we get
\begin{equation}\label{eq4.2}
\textstyle w_h(\alpha)\ge(\ell-k+2)\left(1-\frac{1}{(q_1-1)(q_2-1)}\right)
\end{equation}
and
\begin{equation}\label{eq4.3}
\textstyle (\ell-k+1)\left(1+\frac{1}{q_1-1}\right)\ge(\ell-k+1)w_h(r_k).
\end{equation}
It is an elementary calculation to check that
\begin{equation}\label{eq4.4}
\textstyle
 (x+1)\left(1-\frac{1}{(q_1-1)(q_2-1)}\right)\ge x\left(1+\frac{1}{q_1-1}\right)~\iff~
  x\le q_1-1-\frac{q_1}{q_2}\,.
\end{equation}
Recall that $\ell\le q_1\le q_2$. If $k\ge 3$,
or if $k=2$ and $\ell<q_1$, then $\ell-k+1\le q_1-1-\frac{q_1}{q_2}$.
In both cases, we can apply Formula (\ref{eq4.4}) to (\ref{eq4.2}) and (\ref{eq4.3}),
with $x = \ell-k+1$,  and obtain
$w_h(\alpha)\ge(\ell-k+1)w_h(r_k)$.
\qed

\begin{proposition}\label{prop4.2}
Let $A$
and ${\bf c}_1\in C_1$, $\cdots$, ${\bf c}_k\in C_k$ be as in (\ref{eq4.1})
and assume that ${\bf c}_k\ne {\bf 0}$.
If one of the following two conditions holds:
\begin{itemize}
\item $k\ge 3$,

\item $k=2$ and $\ell<q_1$,
\end{itemize}
then
$$w_{h}\big(({\bf c}_1,\cdots,{\bf c}_k,{\bf 0},\cdots,{\bf 0})A\big)
  \ge(\ell-k+1)w_h({\bf c}_k).$$
\end{proposition}

\noindent{\bf Proof.}~
Let $c_{i_1k},\cdots,c_{i_wk}$ be all the non-zero entries of
${\bf c}_k=(c_{1k},\cdots,c_{nk})^T$.
Then $w_h({\bf c}_k)=w_h(c_{i_1k})+\cdots+w_h(c_{i_wk})$.
Noting that the $i$th row of the matrix
$({\bf c}_1,\cdots,{\bf c}_k,{\bf 0},\cdots,{\bf 0})A$ is
$c_{i1}A_1+\cdots+c_{ik}A_k$, where $A_1$, $\cdots$, $A_k$ are as above,
we have
$$\begin{array}{lcl}
w_{h}\big(({\bf c}_1,\cdots,{\bf c}_k,{\bf 0},\cdots,{\bf 0})A\big)
&=&\sum_{i=1}^n w_h\big(c_{i1}A_1+\cdots+c_{ik}A_k\big)\\[5pt]
&\ge&\sum_{t=1}^w w_h\big(c_{i_t1}A_1+\cdots+c_{i_tk}A_k\big)\\[5pt]
&\ge&\sum_{t=1}^w (\ell-k+1)w_h(c_{i_tk})\\[5pt]
&=&(\ell-k+1)w_h({\bf c}_k),
\end{array}$$
where the second ``$\ge$'' follows from Lemma \ref{lem4.1}.
\qed

\subsection{$2\times q_1$ Non-singular by Columns Matrices}
In the following, we let $q=q_1$ and assume that
$A=\begin{pmatrix}1&1&\cdots&1\\ \beta_1&\beta_2&\cdots&\beta_q\end{pmatrix}$
is a $2\times q$ matrix over $R$ which is non-singular by columns.
Write
$$
 \beta_j=(u_{1j},u_{2j},\cdots,u_{sj})\in R_1\times R_2\times\cdots\times R_s.
$$
Let $a,b\in R$ with $b\ne 0$, and write $a=(a_1,\cdots,a_s)$ and
$b=(b_1,\cdots,b_s)$ with $a_t,b_t\in R_t$ for $t=1,\cdots,s$.
Consider the word
\begin{equation}\label{eq4.5}
\alpha=(a,b)A=(\alpha_1,\cdots,\alpha_q)\in R^q,
\end{equation}
where
$$
 \alpha_j=a+b\beta_j=\big(a_1+b_1u_{1j},\;\cdots,\;a_s+b_su_{sj}\big),
 \qquad j=1,\cdots,q.
$$
Then $w_H(\alpha)\ge q-1$.
From Lemma \ref{lem4.1}(i), we have seen that
\begin{equation}\label{eq4.6}
 w_h(\alpha) = (q-1)w_h(b)\, \qquad\mbox{if~~} w_H(\alpha)=q-1.
\end{equation}
In the following, we further assume that $\alpha_j\ne 0$ for all $j=1,\cdots,q$.

\begin{lemma}\label{lem4.3}
 If $w_h(b)=1$, then $w_h(\alpha)\ge(q-1)w_h(b)$.
\end{lemma}

\noindent{\bf Proof.}~
Since $w_h(b)=1$, there is at least one $k$ such that $b_k\notin J_k^{e_k-1}$. Take
$I_k=R_1\times\cdots\times R_{k-1}\times J_k^{e_k-1}\times R_{k+1}\times\cdots\times R_s$,
and consider the quotient ring $\bar R_k:=R/I_k\cong R_k/J_k^{e_k-1}$.
Then the matrix $\bar A$ over $\bar R_k$
is still non-singular by columns (see Lemma \ref{lemma 2.1}),
$\bar b=\bar b_k\ne 0$, and the
elements
$$
 \alpha_j=\big(a_1+b_1u_{1j},\;\cdots,\;
   a_k+b_ku_{kj},\;\cdots,\;a_s+b_s u_{sj}\big),\quad j=1,\cdots,q,
$$
are mapped to
$$
 \bar\alpha_j=\bar a_k+\bar b_k\bar u_{kj},\qquad j=1,\cdots,q.
$$
Then, for the word
$\bar\alpha=(\bar a,\bar b)\bar A=\big(\bar\alpha_1,\cdots,\bar\alpha_q\big)$
over $\bar R_k$, its Hamming weight satisfies $w_H(\bar\alpha)\ge q-1$. Since $q\ge 2$,
there is at least one non-zero entry, say $\bar\alpha_t\ne 0$,
i.e., $a_k+b_k u_{kt}\notin J_k^{e_k-1}$. Hence, $w_h(\alpha_t)=1$.
Noting that $w_h(\alpha_j)\ge 1-\frac{1}{(q-1)(q_2-1)}$
for $j\ne t$ (see Formula (\ref{eq2.8})), we have
$$\textstyle
 w_h(\alpha)=\sum\limits_{j=1}^qw_h(\alpha_j)\ge 1+(q-1)\left(1-\frac{1}{(q-1)(q_2-1)}\right)
 \ge q-1=(q-1)w_h(b). \eqno\Box
$$

Note that $w_h(b)\ne 1$ if and only if $b_t\in J_t^{e_t-1}$ for $t=1,\cdots,s$.

\begin{lemma}\label{lem4.4}
 If $b=(b_1,\cdots,b_s)$ with $b_t\in J_t^{e_t-1}$, for $t=1,\cdots,s$,
 and $w_h(a)=1$, then $w_h(\alpha)\ge(q-1)w_h(b)$.
\end{lemma}

\noindent{\bf Proof.}~
Similar to the proof above, we can assume that $a_k\notin J_k^{e_k-1}$
for some $k$. Since $b_k\in J_k^{e_k-1}$,
it follows that $a_k+b_k u_{kj}\notin J_k^{e_k-1}$ for all $j=1,\cdots,q$.
Thus $w_h(\alpha_j)=1$ for all $j=1,\cdots,q$, and
$$\textstyle
 w_h(\alpha)=\sum\limits_{j=1}^qw_h(\alpha_j)=q=(q-1)\left(1+\frac{1}{q-1}\right)
 \ge (q-1)w_h(b). \eqno\Box
$$

\medskip From now on, we further assume that
\begin{equation}\label{eq4.7}
a=(a_1,\cdots,a_s),~ b=(b_1,\cdots,b_s)~~
 \mbox{with}~ a_t,b_t\in J_t^{e_t-1}~~\mbox{for}~ t=1,\cdots,s,
 \end{equation}
and let
\begin{equation}\label{eq4.8}
 T_a=\{1\le t\le s\mid a_t\ne 0\}, \quad T_b=\{1\le t\le s\mid b_t\ne 0\}, \quad
   T=T_a\cup T_b.
\end{equation}

\begin{lemma}\label{lem4.5}
Let $t_0=\min\limits_{t\in T}\,t$.
If $q<q_{t_0}$, then $w_h(\alpha)\ge(q-1)w_h(b)$.
\end{lemma}

\noindent{\bf Proof.}~ Since $b=(0,\cdots,0,b_{t_0},\cdots,b_s)$,
by Formula (\ref{eq2.7}), we have that $w_h(b)\le 1+\frac{1}{q_{{t_0}}-1}$.
On the other hand, $a_j+b_j u_{tj}=0$ for any $t<t_0$, so
$$\alpha_j=(0,\cdots,0,a_j+b_j u_{t_0j},\cdots,a_j+b_j u_{sj}),$$
hence $w_h(\alpha_j)\ge 1-\frac{1}{(q_{t_0}-1)(q_{t_0+1}-1)}$ (when $t_0 =s$, set
$q_{t_0+1}$ to be any integer greater than $q_{t_0}$).
Since $q-1\le q_{t_0}-2\le q_{t_0}-1-\frac{q_{t_0}}{q_{t_0+1}}$,
we can use (\ref{eq4.4})
to obtain
$$\begin{array}[b]{rcl}
 w_h(\alpha)&=&\textstyle\sum\limits_{j=1}^q w_h(\alpha_j)\ge
  q\left(1-\frac{1}{(q_{t_0}-1)(q_{t_0+1}-1)}\right)\\
  &\ge&\textstyle (q-1)\left(1+\frac{1}{q_{{t_0}}-1}\right)\ge(q-1)w_h(b).
\end{array}\eqno\Box$$

\medskip In the following, we further assume that

\begin{equation}\label{eq4.9}
1\in T\,, \quad \mbox{and~ $q_2>q_1+1$ if $s>1$.}
\end{equation}

\begin{lemma}\label{lem4.6}
If $T_b=\{t'\}$ contains only one index $t'$
with $1\le t'\le s$, then $w_h(\alpha)\ge(q-1)w_h(b)$.
\end{lemma}

\noindent{\bf Proof.}~ First, assume that $t'=1$.
Then $b=(b_1,0,\cdots,0)\in R_1\times\cdots\times R_s$
with $0\ne b_1\in J_1^{e_1-1}$, so $w_h(b)=1+\frac{1}{q-1}$ (recall that $q=q_1$),
and
$$
 \alpha_j=\big(a_1+b_1 u_{1j},\;a_2,\;\cdots,\;a_s\big).
$$
Taking $I=J_1\times R_2\times\cdots\times R_s$ and $\bar R=R/I\cong R_1/J_1=F_1$, then
$$\bar A=\begin{pmatrix}1&1&\cdots&1\\
 \bar u_{11}&\bar u_{12}&\cdots&\bar u_{1q}\end{pmatrix}
$$
is a matrix over the field $F_1$ which is still non-singular by columns,
so as elements of the field $F_1$, the entries
$\bar u_{11},\bar u_{12},\cdots,\bar u_{1q}$ must be distinct.
Since $|F_1|=q$, we conclude that $\bar u_{11},\bar u_{12},\cdots,\bar u_{1q}$
must consist of all the elements of $F_1$. By Eqn~(\ref{steps PIR}),
$b_1 u_{11},\;b_1 u_{12},\;\cdots,\;b_1 u_{1q}$ are just all
the elements of $J_1^{e_1-1}$, hence
$$ a_1+b_1 u_{11},\;a_1+b_1 u_{12},\;\cdots,\;a_1+b_1 u_{1q} $$
are again just all the elements of $J_1^{e_1-1}$.
In other words, exactly one of them is $0$,
and the other $(q-1)$ terms are non-zero.
By Formula (\ref{eq2.7}),
$$ w_h(\alpha_j)=\begin{cases}
 1-(-1)^{|T|}\cdot\prod\limits_{t\in T}\frac{1}{q_t-1}\,,
    &\mbox{if $a_1+b_1 u_{1j}\ne 0$,}\\
 1+(-1)^{|T|}\cdot\prod\limits_{1\ne t\in T}\frac{1}{q_t-1}\,,
    & \mbox{if $a_1+b_1 u_{1j}=0$.} \end{cases}$$
Therefore,
\begin{eqnarray*}
 w_h(\alpha)&=&\sum_{j=1}^q w_h(\alpha_j)\\
 &=&\left(1 + (-1)^{|T|}\cdot\prod\limits_{1\ne t\in T}\frac{1}{q_t-1}\right)
  +(q-1)\left(1 - (-1)^{|T|}\cdot\prod\limits_{t\in T}\frac{1}{q_t-1}\right)\\
 &=&1 + (-1)^{|T|}\cdot\prod\limits_{1\ne t\in T}\frac{1}{q_t-1}+(q-1)
   - (-1)^{|T|}\cdot \prod\limits_{1\ne t\in T}\frac{1}{q_t-1}\\
 &=&q=(q-1)\left(1+\frac{1}{q-1}\right)=(q-1)w_h(b).
\end{eqnarray*}
Note that the above argument still works well for $T=\{1\}$
(in particular, it works well for $s=1$)
provided we adopt the convention that
$\prod\limits_{1\ne t\in T}\frac{1}{q_t-1}=1$.

\medskip
Next, we assume that $t'>1$. Then $s\ge 2$ and
$$\textstyle w_h(b)\le 1+\frac{1}{q_2-1},$$
$$\textstyle w_h(\alpha_j)\ge 1-\frac{1}{(q-1)(q_2-1)},\qquad j=1,\cdots,q. $$
Since $q_2\ge q+2$, i.e., $q_2-1\ge q+1$, and $q\ge 2$, it follows that:
$$\begin{array}{rcl}
w_h(\alpha)-(q-1)w_h(b)&\ge& q\left(1-\frac{1}{(q-1)(q_2-1)}\right)
 -(q-1)\left(1+\frac{1}{q_2-1}\right)\\[5pt]
 &=& 1-\frac{q}{(q-1)(q_2-1)}-\frac{q-1}{q_2-1}\\[5pt]
 &=& 1-\frac{q}{(q-1)(q_2-1)}-\frac{(q-1)^2}{(q-1)(q_2-1)}\\[5pt]
 &\ge&1-\frac{q+(q-1)^2}{(q-1)(q+1)}=\frac{q-2}{q^2-1}\ge 0.
\end{array}$$
In other words, $w_h(\alpha)\ge (q-1)w_h(b)$. \qed

\begin{lemma}\label{lem4.7}
If $|T_b|\ge 2$, then $w_h(\alpha)\ge(q-1)w_h(b)$.
\end{lemma}

\noindent{\bf Proof.}~ By Formula (\ref{eq2.7}),
$w_h(b)=1-(-1)^{|T_b|}\prod\limits_{t\in T_b}\frac{1}{q_t-1}$, thus
$$
 w_h(b)\le 1+\frac{1}{(q-1)(q_2-1)(q_3-1)}
$$
(put any $q_3\ge q_2$ if $s=2$). On the other hand, by (\ref{eq2.8}), we have
$$
 w_h(\alpha_j)\ge 1-\frac{1}{(q-1)(q_2-1)}.
$$
Noting that $q_2-1\ge q+1$ and $q_3-1>q-1\ge 1$, we have that
$$\begin{array}{rl}
 & w_h(\alpha)-(q-1)w_h(b)\\
 ~\ge &  q\left(1-\frac{1}{(q-1)(q_2-1)}\right)
 -(q-1)\left(1+\frac{1}{(q-1)(q_2-1)(q_3-1)}\right)\\[5pt]
 ~= &  1-\frac{q}{(q-1)(q_2-1)}-\frac{1}{(q_2-1)(q_3-1)}\\[5pt]
 ~> &  1-\frac{q}{(q-1)(q+1)}-\frac{1}{(q+1)(q-1)}\\[5pt]
 ~= &  1-\frac{1}{q-1}~\ge~0.
\end{array}$$
We have obtained the desired inequality $w_h(\alpha)\ge(q-1)w_h(b)$. \qed

\medskip Summarizing Eqn (\ref{eq4.6}) and Lemmas \ref{lem4.3}--\ref{lem4.7},
we have that, if $q_2>q_1+1$ provided $s>1$,
then the homogeneous weight of the word (\ref{eq4.5}) satisfies
\begin{equation}\label{eq4.10}
 w_h\big((a,b)A\big)\ge(q-1)w_h(b).
\end{equation}
Thus, similar to Proposition \ref{prop4.2},
we obtain the following conclusion.

\begin{proposition}\label{prop4.8}
Let $A=(a_{ij})_{m\times q_1}$ be non-singular by columns, and let
${\bf c}_1\in C_1$, ${\bf c}_2\in C_2$ and ${\bf c}_2\ne {\bf 0}$.
Assume the following condition holds
\begin{itemize}
\item $q_2>q_1+1$ provided $s>1$.
\end{itemize}
Then
$$
 w_{h}\big(({\bf c}_1,\,{\bf c}_2,{\bf 0},\cdots,{\bf 0})A\big)
  \ge(q_1-1)w_h({\bf c}_2).
$$
\end{proposition}

\noindent{\bf Proof.}~
It is clear that, for any $q_1\times q_1$ diagonal matrix
$D$ whose diagonal entries are all units of $R$, we have that
\begin{equation}\label{eq4.11}
 w_{h}\big(({\bf c}_1,\,{\bf c}_2,{\bf 0},\cdots,{\bf 0})AD\big)
 =w_{h}\big(({\bf c}_1,\,{\bf c}_2,{\bf 0},\cdots,{\bf 0})A\big).
\end{equation}
Since $A$ is non-singular by columns,
any element of the first row of $A$ is a unit of $R$,
so there is a suitable diagonal matrix $D$ such that
all entries of the first row of $AD$ are $1$.
Thus, we can assume that the first row of $A$ is the all-$1$ vector.
Then, as in the proof of Proposition \ref{prop4.2},
we can obtain the conclusion of the proposition by using (\ref{eq4.10}) .
\qed

\subsection{$1\times \ell$ Non-singular by Columns Matrices}

A $1 \times\ell$ non-singular by columns matrix is none
other than a matrix consisting of only one row,
all of whose entries are units. This is essentially the key ingredient
in the proof of the following result.

\begin{proposition}\label{prop4.9}
Let $A=(a_{ij})_{m\times\ell}$ be non-singular by columns,
and ${\bf 0}\ne {\bf c}_1=(c_{11},\cdots,c_{n1})^T\in C_1$. Then
$$w_{h}\big(({\bf c}_1,\,{\bf 0},\cdots,{\bf 0})A\big)\ge\ell w_h({\bf c}_1).$$
\end{proposition}

\noindent{\bf Proof.}~ By the non-singularity by columns of $A$,
all the entries $a_{11},\cdots,a_{1\ell}$ of the first row of $A$
are units in $R$. Thus,
$$
w_{h}\big(({\bf c}_1,\,{\bf 0},\cdots,{\bf 0})A\big)=
\sum_{j=1}^\ell w_h(a_{1j}{\bf c}_1)
=\sum_{j=1}^{\ell}w_h({\bf c}_1)=\ell w_h({\bf c}_1). \eqno\Box
$$

\subsection{Completion of the Proof of Theorem \ref{Theorem}}

Now we can complete the proof of Theorem \ref{Theorem}.

First, we prove Inequality (\ref{main ineq}).

Let ${\bf c}=({\bf c}_1,\cdots,{\bf c}_m)A$ and
${\bf c'}=({\bf c}'_1,\cdots,{\bf c}'_m)A$ be any two distinct codewords
of the code $C$. Then, not all of  ${\bf b}_j = {\bf c}_j-{\bf c}'_j$, for $j=1,\cdots,m$,
are zero.
Hence, ${\bf c}-{\bf c'}=({\bf b}_1,\cdots,{\bf b}_m)A\ne{\bf 0}$ and
$$
d_h({\bf c},{\bf c'})=w_h({\bf c}-{\bf c'})
=w_h\big(({\bf b}_1,\cdots,{\bf b}_m)A\big).
$$
It is enough to show that $d_h({\bf c},{\bf c'})$
is bounded below by one of the entries in the braces
of the right hand side of (\ref{main ineq}) of Theorem \ref{Theorem}.
Since not all of ${\bf b}_1,\cdots,{\bf b}_m$ are ${\bf 0}$,
there is an index $k$ with $1\le k\le m$ such that
${\bf b}_k\ne{\bf 0}$ but ${\bf b}_{k+1}=\cdots={\bf b}_{m}={\bf 0}$.

If $k=1$, by Proposition \ref{prop4.9}, we have
$$
 d_h({\bf c},{\bf c'})=w_h\big(({\bf b}_1,{\bf 0},\cdots,{\bf 0})A\big)
 \ge \ell w_{h}({\bf b}_1)=\ell w_h({\bf c}_1-{\bf c}'_1)\ge \ell d_h(C_1).
$$

Suppose that $k=2$,  then $m\ge 2$ and, by Lemma \ref{lemma 2.2}, $\ell\le q_1$.

If $\ell<q_1$, by Proposition \ref{prop4.2}, we have
$$\begin{array}{ccl}
 d_h({\bf c},{\bf c'})
 &=&w_h\big(({\bf b}_1,{\bf b}_2,{\bf 0},\cdots,{\bf 0})A\big)
  \ge (\ell-1)w_{h}({\bf b}_2)\\[3pt]
 &=&(\ell-1)w_h({\bf c}_2-{\bf c}'_2)\ge(\ell-1)d_h(C_2).
\end{array}$$
Otherwise, $\ell=q_1$, and by Proposition \ref{prop4.8}, we still have
$$\begin{array}{ccl}
 d_h({\bf c},{\bf c'})
 =w_h\big(({\bf b}_1,{\bf b}_2,{\bf 0},\cdots,{\bf 0})A\big)
  \ge (\ell-1)w_{h}({\bf b}_2)\ge(\ell-1)d_h(C_2).
\end{array}$$

The remaining case is that of $k>2$. By Proposition \ref{prop4.2}, we obtain
$$\begin{array}{ccl}
 d_h({\bf c},{\bf c'})
 &=&w_h\big(({\bf b}_1,\cdots,{\bf b}_k,{\bf 0},\cdots,{\bf 0})A\big)
  \ge (\ell-k+1)w_{h}({\bf b}_k)\\[3pt]
 &=&(\ell-k+1)w_h({\bf c}_k-{\bf c}'_k)\ge(\ell-k+1)d_h(C_k).
\end{array}$$

\medskip
Next, assume that $A$ is column-permutably upper triangular.
Since any permutation of columns does not change
the weights and other parameters of the
resulting codewords, we can assume that $A$ is upper triangular:
$$
 A=\begin{pmatrix}a_{11}&a_{12}&\cdots&a_{1m}&\cdots&a_{1\ell}\\
    &a_{22}&\cdots&a_{2m}&\cdots&a_{2\ell}\\ &&\ddots&\vdots&\vdots&\vdots\\
    &&& a_{mm}&\cdots&a_{m\ell} \end{pmatrix}.
$$
Since $A$ is non-singular by columns,
every element of the first row is a unit of~$R$.
Similarly, every $(2\times 2)$-determinant within the first two rows is a unit,
in particular, $\det\begin{pmatrix}a_{11}&a_{1j}\\ &a_{2j}\end{pmatrix}$
is a unit, i.e., every $a_{2j}$, for $j=2,\cdots,\ell$, is a unit of~$R$.
Continuing this reasoning, we see that
\begin{itemize}
\item all $a_{ij}$ for $1\le i\le m$ and $i\le j\le \ell$ are units of $R$.
\end{itemize}
\noindent
For any $k$ with $1\le k\le m$, take ${\bf c}_k,{\bf c}'_k\in C_k$ such that
$d_h({\bf c}_k,{\bf c}'_k)=d_h(C_k)$. We have two codewords of $C$ as follows:
$$
 {\bf c}=({\bf 0},\cdots,{\bf 0},{\bf c}_k,{\bf 0},\cdots,{\bf 0})A,\qquad
 {\bf c}'=({\bf 0},\cdots,{\bf 0},{\bf c}'_k,{\bf 0},\cdots,{\bf 0})A,
$$
whose homogeneous distance is
\begin{eqnarray*}
 d_h({\bf c},{\bf c}')&=&w_h({\bf c}-{\bf c}')
 =w_h\big(({\bf 0},\cdots,{\bf 0},{\bf c}_k-{\bf c}'_k,
       {\bf 0},\cdots,{\bf 0})A\big)\\
 &=&w_h\big({\bf 0},\cdots,{\bf 0},a_{kk}({\bf c}_k-{\bf c}'_k),
       \cdots,a_{k\ell}({\bf c}_k-{\bf c}'_k)\big)\\
 &=&\sum_{j=k}^\ell w_h\big(a_{kj}({\bf c}_k-{\bf c}'_k)\big)
   =\sum_{j=k}^\ell w_h({\bf c}_k-{\bf c}'_k)\\
 &=&(\ell-k+1)d_h(C_k).
\end{eqnarray*}
Thus $d_h(C)\le\min\{\ell d_h(C_1),\cdots,(\ell-m+1)d_h(C_m)\}$.
It follows that equality must hold in (\ref{main ineq}).

\medskip
Finally, assume that $C_1,\cdots,C_m$ are linear
and $C_1\supseteq\cdots\supseteq C_m$.
Write $A=(a_{ij})_{m\times \ell}$. Since $a_{11}$ is a unit of $R$,
we can add a suitable multiple of the first row to the $i$th row,
for each $2\le i\le m$, such that the entries
of the first column of $A$ below $a_{11}$ are changed into $0$, that is,
there are $b_{21},\cdots,b_{m1}\in R$ such that
$$
  \begin{pmatrix}1\\ b_{21}&1\\ \vdots&&\ddots\\ b_{m1}&&&1\end{pmatrix}
  \begin{pmatrix}a_{11}&a_{12}&\cdots&a_{1\ell}\\a_{21}&a_{22}&\cdots&a_{2\ell}\\
   \vdots&\vdots&\vdots&\vdots\\ a_{m1}&a_{m2}&\cdots&a_{m\ell}\end{pmatrix}
  =\begin{pmatrix} a_{11}&a_{12}&\cdots&a_{1\ell}\\ &a'_{22}&\cdots&a'_{2\ell}\\
   &\vdots&\vdots&\vdots\\ &a'_{m2}&\cdots&a'_{m\ell}\end{pmatrix}.
$$
Similarly, $a'_{22}$ is also a unit of $R$, and we can
add a suitable multiple of the second row to the $i$th row,
for $3\le i\le m$, such that the entries below $a'_{22}$
of the second column are changed into $0$. Continuing in the same
manner, we obtain a lower triangular $m\times m$ matrix
$$P=\begin{pmatrix}1\\ b_{21}&1\\ \vdots&\vdots&\ddots\\
   b_{m1}&b_{m2}&\cdots&1\end{pmatrix}$$
such that $PA$ is an upper triangular matrix,
which is still non-singular by columns.

Since
\begin{itemize}
\item $ C=[C_1,\cdots,C_m]A=\big([C_1,\cdots,C_m]P^{-1}\big)(PA)$,
\item $P^{-1}$ still has the form
$P^{-1}=\begin{pmatrix}1\\ b'_{21}&1\\ \vdots&\vdots&\ddots\\
   b'_{m1}&b'_{m2}&\cdots&1\end{pmatrix},$
\item $[C_1,\cdots,C_m]P^{-1}=[C_1,\cdots,C_m]$
(since $C_1,\cdots,C_m$ are linear and
$C_1\supseteq C_2\supseteq\cdots\supseteq C_m$),
\end{itemize}
it follows that
$$
 C=[C_1,\cdots,C_m](PA),
$$
where $PA$ is upper triangular. Hence, by the result above,
equality holds in~(\ref{main ineq}). \qed

\section*{Acknowledgements}
Quite a part of this work was done while the first and third authors were
visiting the Division of Mathematical Sciences,
School of Physical and Mathematical Sciences,
Nanyang Technological University, Singapore, in Autumn 2011.
They are grateful for the hospitality and support.
They also thank NSFC for the support through
Grants No.~11271005 and No.~11171370.
Part of the work was also done when the second author
was visiting the School of Mathematics and Statistics,
Central China Normal University, Wuhan, in Spring 2012.
The author acknowledges the support and hospitality received.
The work of this author was partially supported
by Singapore National Research Foundation
Competitive Research Programme NRF-CRP2-2007-03.

\end{document}